\documentclass{article}

\def\appname{\textit{EvoChain}}

\usepackage{arxiv}

\usepackage[utf8]{inputenc} 
\usepackage[T1]{fontenc}    
\usepackage{hyperref}       
\usepackage{url}            
\usepackage{booktabs}       
\usepackage{amsfonts}       
\usepackage{nicefrac}       
\usepackage{microtype}      
\usepackage{lipsum}
\usepackage{graphicx}
\graphicspath{ {./images/} }

\usepackage{algorithmicx}
\usepackage{textcomp}
\usepackage{xcolor}
\usepackage{comment}
\usepackage{cleveref}
\usepackage[ruled, vlined, linesnumbered, norelsize]{algorithm2e}
\usepackage{algpseudocode}
\usepackage{pgfplots}
\usepackage{pgfplotstable}
\usepackage{caption}
\usepackage{subcaption}
\usepackage{todonotes}
\pgfplotsset{compat=1.17}
\usepackage{acronym}

\title{EvoChain: A Recovery Approach for\\Permissioned Blockchain Applications}

\author{
 Francisco Faria \\
  INESC-ID, Instituto Superior T\'{e}cnico,\\ Universidade de Lisboa \\
  \texttt{francisco.faria@inesc-id.pt} \\
   \And
 Samih Eisa \\
  INESC-ID, Instituto Superior T\'{e}cnico,\\ Universidade de Lisboa  \\
  \texttt{samih.eisa@inesc-id.pt} \\
   \And
 David R. Matos \\
  INESC-ID, Instituto Superior T\'{e}cnico,\\ Universidade de Lisboa \\
  \texttt{david.r.matos@tecnico.ulisboa.pt} \\
  \And
 Miguel L. Pardal \\
  INESC-ID, Instituto Superior T\'{e}cnico,\\ Universidade de Lisboa \\
  \texttt{miguel.pardal@tecnico.ulisboa.pt} \\
}

\begin{document}

\maketitle

\begin{abstract}
Blockchain technology supports decentralized, consensus-driven data storage and processing, ensuring integrity and auditability. It is increasingly adopted for use cases with multiple stakeholders with shared ownership scenarios like digital identity and supply chain management. However, real-world deployments face challenges with mistakes and intrusions.

This article presents \appname{}, a chaincode framework extension introducing \emph{controlled mutability} for data redaction and recovery under time-limited or specific conditions. This mechanism allows corrections during a grace period before immutability takes effect.

We validated our approach using WineTracker, a Hyperledger Fabric-based supply chain application. It enables some users to cancel unwanted operations while preserving the blockchain security and maintaining data consistency. Performance evaluations showed minimal overhead with functional benefits.

\end{abstract}


\section{Introduction}
\label{chap:intro}

Blockchain technology has been a hot topic of research since the introduction and subsequent success of the Bitcoin cryptocurrency~\cite{NakamotoBitcoin}. Blockchain introduces the possibility of maintaining a distributed ledger of transactions without a central authority.
Instead, the ledger is maintained by a peer-to-peer network that stores data and uses cryptographic primitives to link data blocks and transactions, allowing the verification of authenticity, integrity, and non-repudiation of transactions. 
Each network peer replicates the ledger and, with the aid of a consensus protocol, agrees on a unified view of the chain. 
These characteristics combine to make an immutable and tamper-resistant system that simplifies data auditing processes.

Immutability is a key blockchain property, but it can also be a hindrance to the adoption of the technology~\cite{biswas_analysis_barriers_2019, supply_chain_blockchain_analysis}. Many practical applications require some flexibility for data redaction~\cite{supply_chain_blockchain_analysis}, to rectify errors made by human or machine actors. Human errors are unavoidable, and intentional misconduct must also be taken into account. Since the nature of the blockchain does not allow for data modification, fraudulent records would persist in the system. These issues can become an obstacle to the adoption of the technology in several industries that require some flexibility, such as finance, insurance, healthcare, and supply chains.

Several approaches have been proposed to cope with the limitations presented and introduce recovery mechanisms for blockchain-based applications~\cite{Politou2021BlockchainMutability,zhang_exploring_2021}. These approaches aim to implement redaction mechanisms on blockchain systems without breaking their core security properties. However, they either require developers to change the underlying blockchain or focus on the recovery of tokens and not arbitrary applications. 
In Section~\ref{subsec:blockchain_immutability}, we provide an analysis of state-of-the-art mutable blockchain systems, where we discuss their vulnerabilities and limitations.

In this work, we propose a new framework, \appname{}, an \textit{evo}lution of block\textit{chain}, that introduces apparent transaction mutability while preserving fundamental block\-chain properties such as integrity, authenticity, and non-repudiation, and maintaining a simple auditing process. \appname{} takes advantage of Write-Ahead Logging (WAL)~\cite{mohan1992aries}, combined with a high-level transactional model, to generate a consistent and mutable view of the chain. Unlike other recovery approaches, \appname{} does not require modifications to the underlying blockchain, since it implements a new data architecture within the Chaincode.
We take a supply chain use case~\cite{supply_chain_blockchain_analysis} to implement and test an application built with the new controllable mutability framework over a CFT blockchain\footnotemark.

\footnotetext{CFT (Crash Fault Tolerance) ensures system reliability by handling failures where nodes stop functioning without malicious behavior, whereas BFT (Byzantine Fault Tolerance) extends this capability by tolerating arbitrary or malicious faults~\cite{castro1999practical}.}

\section{Background}
\label{chap:back}

This Section provides a background on blockchain technology, describes the main properties of a standard blockchain system, introduces intrusion recovery, and concludes with the state-of-the-art of intrusion recovery systems for blockchain applications.

\subsection{Intrusion Recovery}
Intrusion recovery is the process of identifying and reversing the effects caused by unintended operations in a system. 
Although new security technologies tend to make intrusions more difficult, they are still unavoidable as long as software contains bugs and users are fallible in practice. 
Two common approaches are generally followed when dealing with intrusion recovery and undoing committed transactions: \emph{rollbacks} and \emph{compensations}.
Rollback is based on the idea of reverting all the activity of an application to a point in time prior to an intrusion, a checkpoint.
Compensations~\cite{korth_formal_1993} undo committed or uncommitted transactions that affect the state by applying special-purpose compensating transactions that revert the state changes caused by the unintended operation. 

\subsection{Blockchain}
\label{sec:blockchain}
Blockchain is a distributed ledger system that operates without the need for central authorities or trusted third parties. A network of nodes with computing capabilities maintains and validates transactions through a consensus protocol.
Blockchain networks can be broadly categorized into two types, regarding the nature of participation: permissionless (public) and permissioned (private or consortium). In permissionless blockchains, such as Bitcoin~\cite{NakamotoBitcoin}, anyone can participate in the network without the need for authorization. This open nature gives the system high resilience but also scalability challenges. Inversely, permissioned blockchains only allow the participation of authorized parties, ensuring access control. These types of blockchains are often used in enterprise settings where data confidentiality and governance are desired characteristics.

A transaction is the fundamental unit of activity in the blockchain. Transactions involve either adding new data to the ledger, updating records, or transferring tokens.
A token is a digital asset representing value, utility, or ownership within a blockchain.
Each validated transaction is added to a storage unit, the block. Each block contains a set of validated transactions, along with metadata and a unique identifier.
Agreement and trust between nodes are achieved by performing a consensus protocol.

There are several consensus protocols, each with its own strengths and trade-offs. Proof-of-Work (PoW)~\cite{dwork1992pricing_pow, NakamotoBitcoin, HashcashAndBitcoin}, Proof-of-Stake (PoS)~\cite{vasin2014blackcoin_pos}, Delegated Proof-of-Stake (DPoS)~\cite{ago_dpos_2017} were designed specifically for blockchain systems, while others, such as Practical Byzantine Fault Tolerance (PBFT)~\cite{castro1999practical}, have been adapted from previous research on fault-tolerant systems.

\subsection{Blockchain Immutability}
\label{subsec:blockchain_immutability}
Immutability is a feature of blockchain technology and ensures that a transaction, once executed, cannot be modified or deleted. This property is supported through the use of cryptographic primitives that are the base of the chain and certifies that a transaction, after being accepted on the chain, cannot be removed or changed (mutated). 

The integrity of the data is ensured by the use of the Merkle Tree data structure~\cite{Merkle1987ADS}, along with the collision resistance and block linking derived from the use of cryptographic hash functions~\cite{al2011cryptographic}.
Immutability is particularly desirable in contexts where data integrity and a simplified auditing process is crucial and necessary, such as in financial transactions.
However, this property may contradict several privacy requirements and data protection rights, as well as the adoption of blockchain technologies for a wider range of applications where data requires some changes.
Theoretical and technical implementations of a mutable system that retains the inherent security of a blockchain system are still in their early stages, although several cryptographic and innovative approaches have been emerging recently~\cite{zhang_exploring_2021, Politou2021BlockchainMutability}. These methods typically rely on one of two strategies: circumventing/bypassing immutability or conditionally removing it.
Bypass strategies involve the way data is represented on the chain, the way data is stored, the introduction of new data structures, or the use of a decentralized set of judges. In addition to avoiding heavy cryptographic primitives, they usually introduce new attack vectors to the system.
Regarding removing strategies, we can also define three main techniques based on the concepts of Consensus~\cite{DAO}, Chameleon-Hash~\cite{RedactabaleBlockchainCharmaleonHash} and Meta-Transactions~\cite{MuchainFabric}.
These strategies involve hard forks of the main chain, cryptographic changes to hash functions, or changes to the blockchain data structures. They usually require heavier cryptographic primitives, introduce delays to the system, and often compromise auditability.

\subsection{Hyperledger Project}
\label{subsec:hlf}

Hyperledger~\cite{Hyperledger-all} is a project framework hosted by The Linux Foundation, which offers the guidelines, standards, and tools to develop cross-industry blockchain technologies. The goal is to provide the necessary infrastructure and standards to develop systems and applications for industrial use cases. Hyperledger includes various subprojects, including Fabric~\cite{Hyperledger} and Caliper~\cite{hyperledger_capilar}.

Hyperledger Fabric~\cite{Hyperledger} is an open-source, permissioned distributed ledger platform designed to be used in enterprise contexts. With its highly modular and configurable architecture, it provides easy adaptability to a broad range of use cases in industries such as banking, finance, and supply chains. It is programmable and supports the use of smart contracts (chaincode) in general-purpose programming languages.
Its modular architecture has several main components: the membership service, responsible for the entities in the network; the peer nodes, responsible for the ledger; the chaincode, responsible for the application logic; the ordering service, responsible for ordering transactions and the consensus protocol; and the peer-to-peer (P2P) protocol, responsible for communication between peers.
Fabric uses the concept of \emph{world state} which refers to the latest values of all key-value pairs in the blockchain network.
Unlike traditional blockchains, Fabric separates the concept of world state and transaction history by implementing versioning of assets, which allows managing changes and updating assets over time.
The world state is maintained in a state database, such as LevelDB or CouchDB, alongside the ledger, for efficient query and update operations. The world state enables quick access to the current value of any asset without having to traverse the entire transaction log.
When a key-value pair is updated, a new version is created, preserving a history of changes in the ledger.

\section{EvoChain}
\label{chap:architecture}

We present \appname{}, a development framework for Chaincode that brings in controlled mutability for decentralized applications. Here we introduce an overview of the components, a general sequence model of a redaction, where a submitted transaction is canceled and not considered to the final state of the system, and the template code of the main components.
\appname{} proposes the use of concepts previously applied by techniques such as WAL~\cite{mohan1992aries} to generate a consistent and correct view of blockchain data.
Taking advantage of the Hyperledger Fabric world state, \appname{} maintains individual logs for each processed transaction in a sequential manner.
The main goal is to propose a chaincode development framework that allows for reversal and recovery of operations.

\subsection{Recovery Scenarios}
\label{subsec:recovery_scenarios}
Here are some of the scenarios where \appname{} is intended to work:

\begin{itemize}
    \item \textbf{S1:} Account Theft: by means of phishing attacks, social engineering, or malware/keylogging, a user loses access to an account and cannot recover;
    \item \textbf{S2:} User fault: a user, inadvertently or due to a misunderstanding, issues a transaction that was not meant to be sent out to the blockchain network;
    \item \textbf{S3:} Incorrect Authorization and Access Controls: insufficient safeguards regarding access control and permissions granted to users create the potential risk of unauthorized entry and improper utilization of features;
    \item \textbf{S4:} Smart contract exploitation:   vulnerabilities and bugs in the chaincode, which can result in unintended behavior and security breaches due to unexpected execution paths or incorrect logic.
\end{itemize}

\subsection{Threat Model}
\label{subsec:threat_model}

We assume that an attacker can perform the following operations from the application level:
\begin{itemize}
    \item \textbf{A1:} A mistakenly issued transaction by an authorized user.
    \item \textbf{A2:} Tricking users to issue unwanted transactions.
    \item \textbf{A3:} Steal user private keys and illegally execute transactions.
    \item \textbf{A4:} Exploit smart contract vulnerabilities.
\end{itemize}

Our model focuses on application-level attacks since the Hyperledger Fabric's design and architecture already address several external threats:
Sybil attacks \cite{douceur_sybil_2002} by using permissioned networks with Membership Service Providers to control participant identities; 
Network Partitioning \cite{network_part_decker} since Fabric's channels and endorsement policies ensure that valid transactions are eventually committed;
Eclipse attacks \cite{eclipse_heilman} by presenting the isolation of validating peers' endorsement policies and gossip protocols; and 
Blockchain Reorganization attacks \cite{Gervais2014IsBA} since Fabric offers a strong consistency model and chain validation process, therefore preventing the tampering of past transactions. 

We also do not consider possible flaws in the underlying consensus algorithms, hash functions, and digital signatures, since delving into underlying technologies is out the scope of this work.

\subsection{Design}

\appname{} provides controlled mutability to application logic.
It brings transaction concepts to a higher level by presenting transaction-based storage in the world state instead of individual object storage. 
By recording changes at the transaction-level, \appname{} allows for the control of Fabric world state, which leads to controlled mutability.
The current programming interface remains similar, but \appname{}  introduces a `cancel' (undo) operation.
Instead of logging data into storage before making permanent changes, an \appname{} transaction is written into the chain as a Fabric transaction and also stored in the world state. These transactions are observed in the generation of the consistent view, but are not automatically considered consolidated and immutable.

\appname{} transactions are divided into two types: MT and CT.

\begin{itemize}

\item Mutable-Transactions (MT) are standard blockchain transactions extended to accommodate four fields: submission time, permanent state time, validity, and delay, as well as the objects regarding the application logic.
MTs are issued in a \textit{pending} validity state, with a submission time related to the time they were issued and a default delay.

\item Canceling-Transactions (CT) work as special-purpose transactions that update the state of previously issued \textit{pending} MTs to a \textit{canceled} validity state.
CTs are accepted if they refer to a MT that has not reached the $mutation$ $policy$ specified, namely if the consolidation delay has expired, or a certain condition, for example, the issue of a specific dependent transaction within the system. That cancellation (undo) will affect every transaction that is dependent of that MT, producing a rollback effect. This rollback will revert the changes performed by a transaction that will be noticeable in the view generation.
\end{itemize}

In \appname{}, transactions that are dependent on the outcome of previous transactions have a mutation policy that must expire in a later timeframe.

As an example, consider transactions $A$ and $B$: $B$ is dependent on $A$: $B \rightarrow A$ if both $B$ and $A$ alter the state of the same object and $A$ was previously issued.
$B$ can be consolidated only if $A$ is also consolidated: $C(B) \Rightarrow C(A)$.

The mutation policy will be programmed in the chaincode of the implemented application and will state who and under which conditions a CT can be issued. The submission time works as a version number of a transaction that will be assigned by the chaincode, to every transaction, with the original MT being the first version of each transaction state.

\subsection{Operations}

Clients still use the Fabric API to communicate with the Fabric Network.
The \appname{} chaincode framework offers four generic operations:
$IssueTransaction$, $CancelTransaction$, $GetAsset$, and $GetTransactions$.

$IssueTransaction$ works as a standard operation to submit a transaction to the network. It submits a transaction that is, considered within the application chaincode, a MT. It is verified by the chaincode and, if valid, submitted.

$CancelTransaction$ is issued with the goal of canceling a MT. The mutation policy is verified by the chaincode that accepts the transaction if it is not yet considered consolidated. If accepted, from that moment forward, the view generated by the system changes and the final object stops affecting the canceled MT.

$GetAsset$ is the standard query operation 
of an asset created in the network, returning the most recent and not affected by canceled transactions version of an object, considering the view generated by the chaincode.

$GetTransactions$ is an extended query operation 
of an asset. Returns every pending and consolidated transaction that affects the state of an object.

The query operations are responsible for consolidating pending transactions, e.g. they apply the expired delay.
This means that the transaction views are updated only when needed.
However, this incorporates some additional overhead in query operations, as discussion in Section~\ref{sec:query-eval}.

\subsection{Mutation policies}

Hyperledger Fabric comes with built-in policies that ensure transaction and block validation, as well as resource access control. 
\appname{} leverages these and extends them to be \textit{mutation policies}.
These consist of a set of rules and conditions that define the interplay between Mutable-Transactions (MT) and Canceling-Transactions (CT) so that modifications to the ledger state are accepted or denied.
They also implement access control on CTs.

A \textit{pending} transaction can still be canceled.
A transaction is \textit{pending} if the associated delay has not already expired or if it has not yet been consolidated by a conditional procedure.
Each transaction is associated at the time of execution with a delay, which can be seconds or minutes, depending on the application context.
This delay is the interval between the moment a transaction is issued and the moment it should become immutable.
The delay can be altered by an administrator, as long as the transactions that it is dependent on expire in a shorter timeframe.

There are two ways that a transaction can become \textit{consolidated}: by expiration or by satisfied condition.
The \textit{expiration} is checked when a later MT is issued, or there is a query.
The view generation component checks the transaction that affected the corresponding object and if its delay has expired.
If a checked transaction has an expired delay, its status is updated to consolidated and can no longer be mutated.
The \textit{condition} is checked and can trigger the consolidation of a previously issued transaction that it is dependent on. This relationship is configured within the chaincode and also depends on the application logic.

Let us consider two examples.
One example with expiration is the following:
a product is received at a shipping center and is temporarily accepted, with an expiration time set to 1 hour.
During that hour, if there is an inspection that detects that the package was actually not meant for that destination, then the transaction can still be canceled.
If not, after 1 hour, it becomes immutable.

Another example, with condition, is as follows:
again, a product is received at a shipping center and is temporarily accepted until there is a later transaction confirming the reception of the whole shipment.
Until there is such confirmation, the transaction can still be canceled.
After the reception of the confirmation, the transaction becomes immutable, as the condition was satisfied.

In the second example, the immutability occurs from a later transaction and not by the simple passage of time.
Both examples show how \appname{} can make the blockchain-based system much more flexible for real business situations.

\subsection{Dependency Graph}

\appname{} takes advantage of Fabric's world state, to access the actual value of a state instead of having to traverse the entire transaction log. 

A dependency graph is generated by these transactions to systematically track the relationship and dependencies between these objects.
This graph plays the role of ensuring that the lifecycle of each object can be easily inferred and analyzed by the ViewGeneration component, especially when multiple transactions and multiple actors are involved.

\begin{figure}[!htb]
  \centering
  \includegraphics[width=0.70\columnwidth]{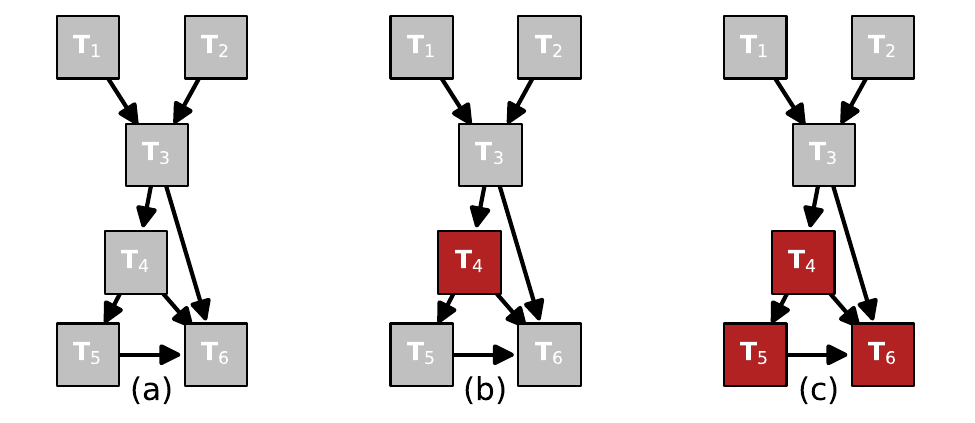}
  \caption[Dependency Graphs]{Dependency Graphs}
  \label{fig:dependency_graph}
\end{figure}

In the graph \ref{fig:dependency_graph}, each node represents a transaction, MT. An MT creates new objects and/or alters the state of existing ones. Multiple actors can issue MT, altering the same objects.
The direct edges between nodes illustrate the dependencies between MT. An edge from T$_1$ to T$_2$ indicates that T$_2$ depends on the previous execution of T$_1$ (T$_i$ < T$_j$, and TS($j$) > TS($i$)). As expressed in Figure~\ref{fig:dependency_graph}, considering a transactional graph (a), if a Mutable Transaction T$_4$ has its state altered (b), it will directly influence the state of T$_5$ and T$_6$ (c).

\subsection{View Generator}

The \appname{} View Generator can generate a specific representation of the data within the system.
This process filters and structures consistent data, using the generated dependency graph, ensuring that users see the last correct and confirmed version of the object based on the programmed and intended criteria. The view generator is responsible for retrieving the latest valid and consolidated object and, from that point, considering pending or consolidated transactions, applying the modifications, and creating the final object that is presented to the application.

The concept of a high-level transactional model is introduced within the chaincode as the data is no longer kept by itself but stored alongside the transaction that affected that object. The components of the transaction, as presented previously, will be considered by the dependency graph when generating the view of the system. 
By integrating data with transactions, \appname{} is able to maintain an interconnected transaction model. 
Therefore, the process of generating a view of the system gains the ability to compose data based on the graph of transactions present in the ledger state.
The state of objects can be modified by multiple transactions, depending on the application context. Therefore, \appname{} uses view generation to traverse the graph and compose, based on the relevant transaction that have been confirmed, are pending, or have been canceled to create a consistent state at the moment of initiating a new transaction within the system.

The pseudocode of the view generation algorithm is presented on Algorithm~\ref{alg:view_generator}.
Every transaction that alters the state of the object is ordered by submission time, leveraging the ordering provided by the ledger.
This allows to view the object state history and undo operations that have been performed. 
When a transaction is issued, it is in a mutable state until the \textit{mutation policy} condition is reached. The transactions that precede that transaction, if not in a permanent state, can alter the new object state within that transaction.

We can guarantee that if a transaction is consolidated and it was issued in a later timeframe than the last canceled transaction's permanent state, that that object is valid and the history can be infered from that object onwards, i.e. it can used as a starting point in the view generation process.

\begin{algorithm}[!htb]
\scriptsize
\caption{View Generation Algorithm}
\label{alg:view_generator}
\KwData{List of transactions that influence object $id$}
\KwResult{Final object}
\SetKwFunction{ViewGenerator}{ViewGenerator}
\SetKwProg{Function}{Function}{:}{end}
\Function{\ViewGenerator{$transactions, id$}}{
    Initialize $initialObject \leftarrow transactions$\;\par
    Initialize $confirmedTransaction$\;\par
    Sort $transactions$ in reverse order $submissionTime$\;\par
    $ConsolidatedTransactions \leftarrow$ $FilterByConsolidated(transactions)$\;

    $CanceledTransactions$ $\leftarrow$ $FilterByCanceled(transactions)$\; 
    
    Sort $CanceledTransaction$ in reverse order $PermanentState$
    
    \ForEach{$consolidatedTransaction$ in $ConsolidatedTransactions$}{
        \If{$confirmedTransaction$ $\neq$ $null$}{
            \textbf{break}\; \tcp*[r]{There are no confirmed transactions}
        }
        
        $Filtered \leftarrow FilterLowerSubTime(CanceledTransactions, $ $consolidatedTransaction)$\;
        
        \If{$Filtered.isEmpty()$} {
            $confirmedTransaction \leftarrow consolidatedTransaction$\;\par
            \textbf{break}\; \tcp*[r]{No canceled transactions, assume the last confirmed}
        }
        
        \ForEach{$canceledTransaction$ in $CanceledTransactions$}{
            \If{$consolidatedTransaction.SubTime > canceledTransaction.PermaState$}{
                $confirmedTransaction \leftarrow consolidatedTransaction$\;\par
                \textbf{break}\; 
                \tcp*[r]{There is a valid confirmed transaction that was not influenced}
               
            }\Else{
                \textbf{break}\; 
                 \tcp*[r]{Confirmed transaction influenced by cancel, check next}
            }
        }
    }

    \If{$confirmedTransaction \neq null$} {
        $initialObject \leftarrow confirmedTransaction.getObject()$\;\par
        $transactions \leftarrow FilterByNotCanceled(transactions)$\;\par
        $transactions \leftarrow FilterByHigherSubTime(transactions, initialObject)$\;
    } \Else{
        $transactions \leftarrow FilterByNotCanceled(transactions)$\;\par
    }
    
    \ForEach{$Transaction$ in $transactions$}{
        $initialObject \leftarrow ApplyChanges(transaction)$\;
    }

    $finalObject \leftarrow initialObject$\;
    
    \Return $finalObject$\;
}
\end{algorithm}

\subsection{Recovery Process Example}

\begin{figure*}[!htb]
    \centering
    \includegraphics[width=1\columnwidth]{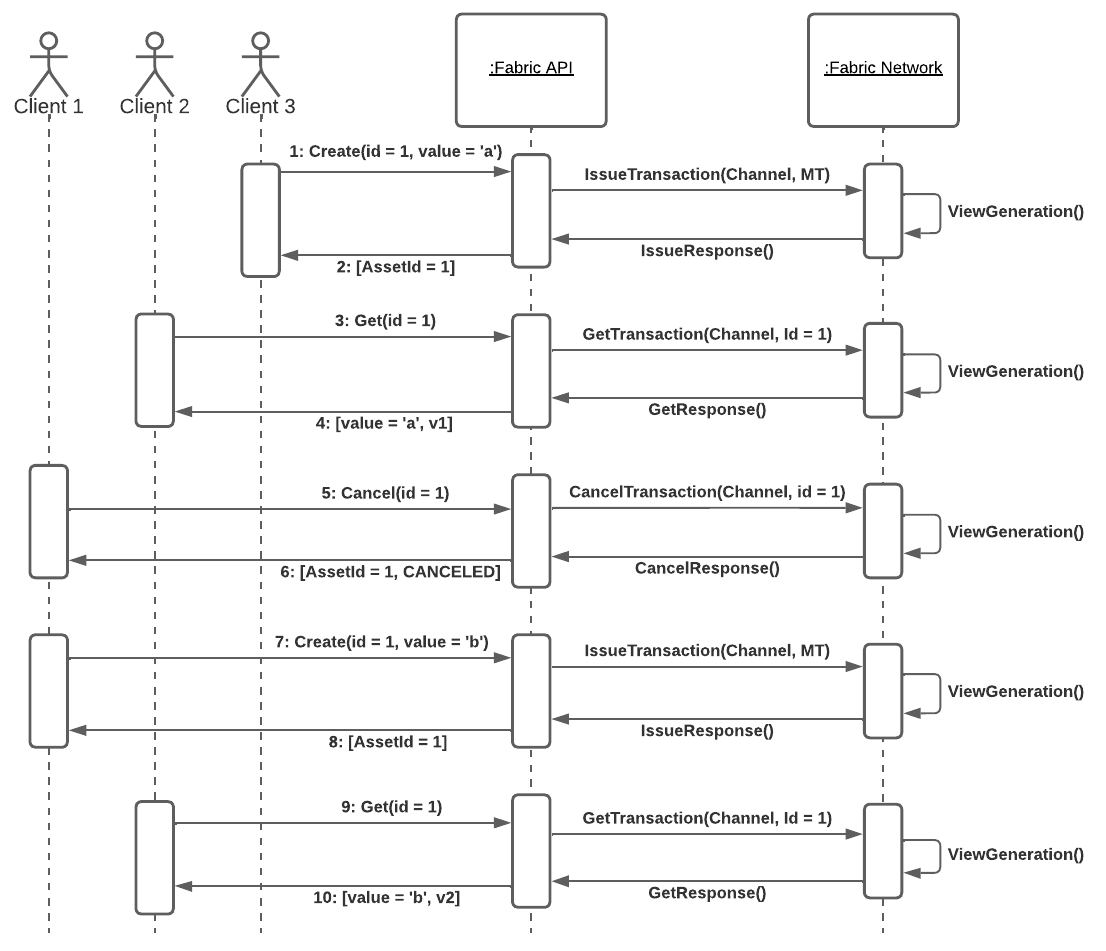}
    \caption[Recovery Example]{Recovery Example}
    \label{fig:recovery_example}
\end{figure*}

A sequence of transactions associated with user operations is shown in Figure~\ref{fig:recovery_example}.
In this example, three clients perform requests through the Fabric API to the Fabric Network. 
Client 1 has permission to both issue and cancel transactions; 
Client 2 is an observer; and 
Client 3 has permission to issue transactions.

\begin{itemize}
    \item Client 3 issues a MT, creating the object with id = 1 and value ``a''. This transaction is, in this case, confirmed by the network. 
    A response containing the properties of the created Asset is forwarded to the client. 
    \item Client 2 queries the ledger with the identifier of the Asset previously created by Client 3. 
    Since now the chaincode handles transaction objects, the system calculates using the ViewGeneration algorithm \ref{alg:view_generator} the Transactions that altered the state of AssetId = 1, generating a consistent view and forwarding the intended object to the client.
    \item Now, Client 1, given that, as explained, as permissions to issue cancel transactions, has permission to cancel certain types of Transactions, issues 
    a CT related to the IssueTransaction 
    of Client 3.
    Assuming that the Transaction issued by Client 3 is not yet consolidated, the validity parameter of the MT issued 
    is altered to \textbf{CANCELED}, and will now be considered as such during the view generation.
    The response to this request 
    is now empty, as the generated view no longer considers the creation of AssetId = 1.
    \item Client 1 now issues a new transaction 
    creating a new AssetId = 1, with value ``b''. This transaction is accepted by the system since, as previously stated, the previous cancel operation changed the new view calculations and the system does not consider the existence of an Asset with identifier 1. 
    The user receives a response with the created asset. 
    \item At last, Client 2 queries the system, 
    looking for the Transactions that affect AssetId = 1. Now, the view generated considers the Asset created by Client 1 and returns 
    the Object created by the same Client, with Id = 1 and value ``b''.
\end{itemize}

Considering Section~\ref{subsec:threat_model}, where the threat model is defined, and taking into account the scenarios presented, \appname{} allows for the recovery of unwanted operations within the system, as long as the mutation policies are properly configured and threats are expected. 

Considering attack scenarios A1 and A2, mistakenly issue of a transaction or tricking user to perform unwanted operations, the ledger state would be recovery by allowing, admins to issue CTs or even allowing users to cancel their own transactions, while on a pending state.

Considering scenarios A3 and A4 where private keys are stolen as user credentials are captured, transactions are executed illegally. With a sufficiently high consolidation delay and the permission for a trusted user (e.g. an administrator) to cancel transactions, a CT after the revoke of authorization of the compromised user, from the network operators, would be sufficient to revert the view of the system to a prior state and mitigate the attack.

\section{Use Case}
To test the \appname{} framework, we created a supply chain network model based on selected models from the relevant literature. The goal was to strengthen the claim that \appname{} can be applied in real-world scenarios.
Our approach was influenced by the developments of Matt Dean \cite {mattdean1_blockchain}, building upon the concept introduced by Biswas et al. \cite{biswas_blockchain_2017}. 
We implemented a simplified model mirroring a Wine Supply Chain.

\subsection{WineTracker}

Biswas et al.~\cite{biswas_blockchain_2017} proposed a blockchain-based Based Wine Supply Chain Traceability System due to the increase in counterfeiting, adulteration, and the use of hazardous chemicals in the wine industry. This industry needed a reliable alternative to current systems, which provide no integrity, where consumers could verify the properties of each batch of wine through the different actors in the supply chain.

\begin{figure*}[!htb]
  \centering
  \includegraphics[width=1\columnwidth]{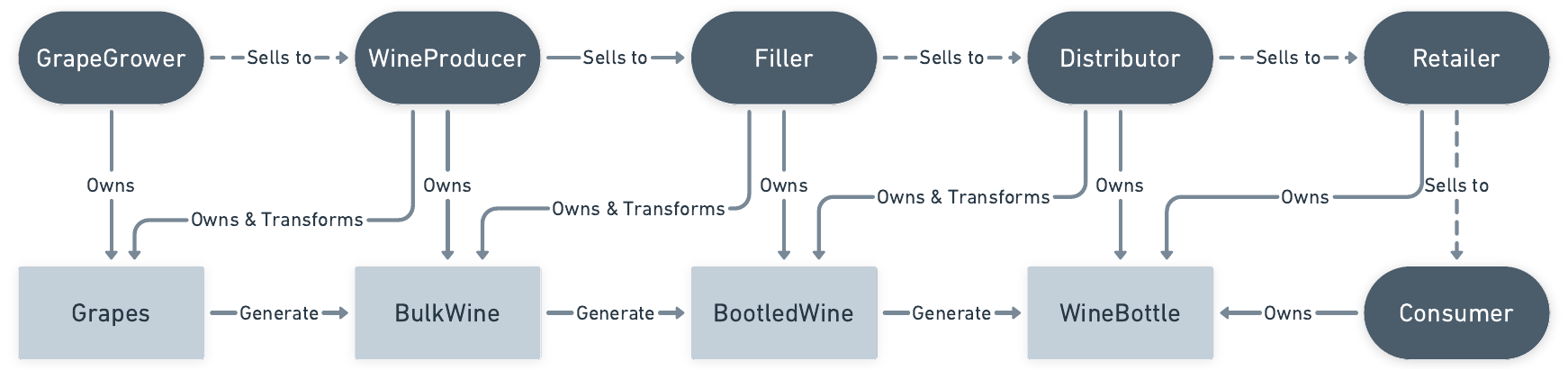}
  \caption[Network Model]{Network Model}
  \label{fig:network_model}
\end{figure*}

WineTracker network models present several entities, each responsible for a part of the supply chain process. Several objects are created, transformed, and sold during this process by a specific actor.
A generic model of the entities is presented in Figure \ref{fig:network_model}.

Growers take no input and create the initial object $Grapes$. This object is sold in quantity to Wine Producers. 
Producers transform and sell it in the form of $BulkWine$ to Fillers.
Fillers fill bottles and associate them with a unique identifier. Finally, these unique bottles are sold to distributors, retailers, and the final consumer.

Assets are submitted to the system as a serialized JSON. Each object has certain attributes inherent to the asset type. In this case, batchId, sellId, owner, and others relevant to the completion of the system.

\subsection{\appname{}WineTracker}

The \appname{}WineTracker is a version of the WineTracker application described before, modified to use \appname{}.
The implementation of the framework mainly converts the data structures where the data is organized, implementing a transactional model within the chaincode to provide control over changes.
Every operation that alters the state of an object is now mapped to a transaction that is issued to the ledger, allowing the implementation of a view generation component that provides flexibility to the end-user by allowing the chaincode itself the possibility of generating a consistent view of the system's data in case of redaction.

Our implementation was developed using a set of standard mutation policies and access controls.
Users in an organization have permission to perform transactions that are related to their role in the network model. The same users and administrators of that organization have permission to issue CTs, administrator can issue CTs to any transaction that has not been consolidated, and users can issue CTs to transactions that have not been consolidated and issued by themselves.
The expired delay was set in the hundreds of seconds, ensuring that the performance tests were only influenced by the consolidation of transactions when expected.

\subsection{\appname{}WineTracker Transactional Architecture}

\label{subsec:evowinetracker_transactional_architecture}
Every ``Transform'' and ``Sell'' operation presented in Fig~\ref{fig:network_model} represents a transaction that can be issued by a certain actor. Previously, these operations changed the state of the objects, submitting that change to the ledger. This change would be submitted to the blockchain and the world state of those same objects would be altered. 
Although it is possible to retrieve from the blockchain the data history for a certain key, it becomes challenging to backtrack the relationship between these changes through the supply chain procedure and ultimately rollback the world state to a previous point in history.
By implementing \appname{} as the basis of WineTracker, the history of objects can be tracked since a transaction relationship exists between them.
The application now has the capability to rollback a process to a previous state.

Every transaction extends the Transaction Class. If that transaction alters the state of an object, it should now contain the new and altered version of that object. Transactions that create objects should contain the initial version of the object created. Simpler transactions that consume or alter the state of an object one single time could only contain the new version of the consumed object, since they do not influence other transactions.
From now on, transactions will be written in the ledger instead of assets, also accompanied by a specific identifier.

\section{Evaluation}
In this Section, we evaluate our implementation of \appname{}. First, the evaluation and the environment setup are described. 
Then, a performance evaluation of our prototype (\appname{}WineTracker), alongside the original version (WineTracker).
\appname{}WineTracker includes extended functionalities in addition to the core functionalities presented in WineTracker, such as the ability to cancel transactions. Both contracts are deployed under identical conditions with the same workload of requests that result in the same output. To conclude, we discuss the results and the limitations.

\subsection{Methodology} 

To assess the capabilities and performance of \appname{}, we took the following steps:

\begin{enumerate}
    \item \textbf{Application implementation}:
    a supply chain application was implemented using Hyperledger Fabric;
    
    \item \textbf{Framework implementation}:
    the \appname{} framework was integrated into the same supply chain application, simulating a real-world use case;
    
    \item \textbf{Performance assessment}:
    comparison of the performance of both versions of the application, measuring request latency (in milliseconds) and throughput (in TPS). This was achieved by performing cloud deployments and collecting metrics;
    
    \item \textbf{Validity assessment}:
    To ensure the validity of the results, a series of functional tests were carried out, focusing on qualitative parameters.
\end{enumerate}

Load tests were conducted to verify the overhead that the modified version of the application using the \appname{} framework brings compared to the original.
The tests focused on measuring latency, throughput, and breakpoints, providing insight into the overhead introduced by integrating \appname{} into an existing application.

Several functionalities were the focus of these tests, since we intended to evaluate,  besides the ``cancel'' operations, the overhead introduced to the core functionalities of the WineTracker, described in Section~\ref{fig:network_model}.

To measure the performance of both the original and modified version, we used Hyperledger Caliper, a benchmarking tool designed to evaluate the performance of blockchain solutions.
The evaluation focused on metrics such as bandwidth, throughput in Transactions per Second (TPS), latency in seconds (s), and mean time to recover (MTTR).

\subsection{Network Topology}

In our Proof of Concept (POC), the blockchain network was designed with four organizations (Org1, Org2, Org3, and Org4), each represented by a single peer and associated with one or more entities within the network model. A single channel facilitates communication and transactions between one and all peers. Four certificate authorities associated with each of the organizations issue certificates to peers and clients within that organization.

\subsection{Configuration}

Our system was deployed on a Google Computer Engine instance, e2-highcpu-16, with 16 vCPUs, 16~GB of memory and a 100~GB Balanced Persistent Disk.
Our tests were performed on Hyperledger Fabric images, version 2.5.0, and the system was deployed using container technology to modularize and manage our system components. Docker v24.0.6 and Docker-Compose 1.29.2 were used to achieve containerization. For benchmarking experiments, we used Hyperledger Caliper v0.5.0.

The following application software and configurations were used:
\begin{itemize}
    \item LevelDB was used as the world state database;
    \item Single application channel, with 4 organizations and 1 peer per organization. Alongside with the orderer, were joined to this channel;
    \item Ordering consensus is achieved by a Raft-based \cite{raft_consensus} consensus algorithm;
    \item Chaincode expressed in the Java programming language using the Contract API was deployed in the network;
    \item Endorsement policy Majority, requiring a majority of endorsing peers to validate a transaction;
    \item Neither private data nor range queries were used;
    \item TLS (Transport Layer Security) was used to secure communication between the entities within the network;
    \item For all other settings, we used the default Fabric policies and configurations. 
\end{itemize}

\subsection{Performance Evaluation}

We evaluated the core functionalities, canceling, and query transactions.

\subsubsection{Core Functionalities}

We evaluated the core functionalities by comparing the two versions in terms of throughput, latency, and memory usage.
Our objective was to assess the overhead of using the version implemented using our framework.

In the first test case (TC1), ten (10) workers were involved in this benchmark, simulating 10,000 transactions per round.
The rate control for each round began at 400 TPS and progressively increased to 1200 TPS.

\begin{figure}[!htb]
    \centering
    \footnotesize
    
    \begin{subfigure}[b]{0.32\textwidth}
        \centering
        \begin{tikzpicture}
            \begin{axis}[
                ybar, bar width=14pt, enlarge x limits=0.8, ylabel={Latency (ms)},
                symbolic x coords={Latency}, xtick=data, ytick={0,10,20,30,40}, ymin=0, ymax=45,
                nodes near coords, nodes near coords style={font=\scriptsize}, every node near coord/.append style={yshift=0.5em},
                width=\textwidth, height=5cm, title={Average Latency}, ylabel near ticks,
                legend style={at={(0.5,-0.25)}, anchor=north, legend columns=-1},
                legend image post style={scale=0.7}
            ]
            \addplot[fill=blue!60, bar shift=-12pt] coordinates {(Latency, 30.19)};
            \addplot[fill=orange!60, bar shift=12pt] coordinates {(Latency, 35.1)};
            \end{axis}
        \end{tikzpicture}
    \end{subfigure}
    \hfill
    \begin{subfigure}[b]{0.32\textwidth}
        \centering
        \begin{tikzpicture}
            \begin{axis}[
                ybar, bar width=14pt, enlarge x limits=0.5, ylabel={Throughput (TPS)},
                symbolic x coords={Throughput}, xtick=data, ytick={0,50,100,150,200}, ymin=0, ymax=220,
                nodes near coords, nodes near coords style={font=\scriptsize}, every node near coord/.append style={yshift=0.5em},
                width=\textwidth, height=5cm, title={Throughput}, ylabel near ticks,
                legend style={at={(0.5,-0.25)}, anchor=north, legend columns=-1},
                legend image post style={scale=0.7}
            ]
            \addplot[fill=blue!60, bar shift=-12pt] coordinates {(Throughput, 180.67)};
            \addplot[fill=orange!60, bar shift=12pt] coordinates {(Throughput, 163.8)};
            \end{axis}
        \end{tikzpicture}
    \end{subfigure}
    \hfill
    \begin{subfigure}[b]{0.32\textwidth}
        \centering
        \begin{tikzpicture}
            \begin{axis}[
                ybar, bar width=14pt, enlarge x limits=0.2, ylabel={Memory (MB)},
                symbolic x coords={Memory}, xtick=data, ytick={0,100,200,300,400,500}, ymin=0, ymax=570,
                nodes near coords, nodes near coords style={font=\scriptsize}, every node near coord/.append style={yshift=0.5em},
                width=\textwidth, height=5cm, title={Average Memory Usage}, ylabel near ticks,
                legend style={at={(0.5,-0.25)}, anchor=north, legend columns=-1},
                legend image post style={scale=0.7}
            ]
            \addplot[fill=blue!60, bar shift=-12pt] coordinates {(Memory, 459.4)};
            \addplot[fill=orange!60, bar shift=12pt] coordinates {(Memory, 469.5)};
            \end{axis}
        \end{tikzpicture}
    \end{subfigure}

    \vspace{0.2cm}
    \begin{tikzpicture}
        \begin{axis}[
            hide axis, xmin=0, xmax=1, ymin=0, ymax=1,
            legend style={draw=none, at={(0.5,-0.1)}, anchor=north, legend columns=-1, /tikz/every even column/.append style={column sep=1cm}}
        ]
        \addlegendimage{fill=blue!60, area legend}
        \addlegendentry{Vanilla}
        \addlegendimage{fill=orange!60, area legend}
        \addlegendentry{EvoChain}
        \end{axis}
    \end{tikzpicture}
    
    \caption{Comparison of Vanilla and EvoChain Versions of WineTracker}
    \label{fig:winetracker_comparison}
\end{figure}
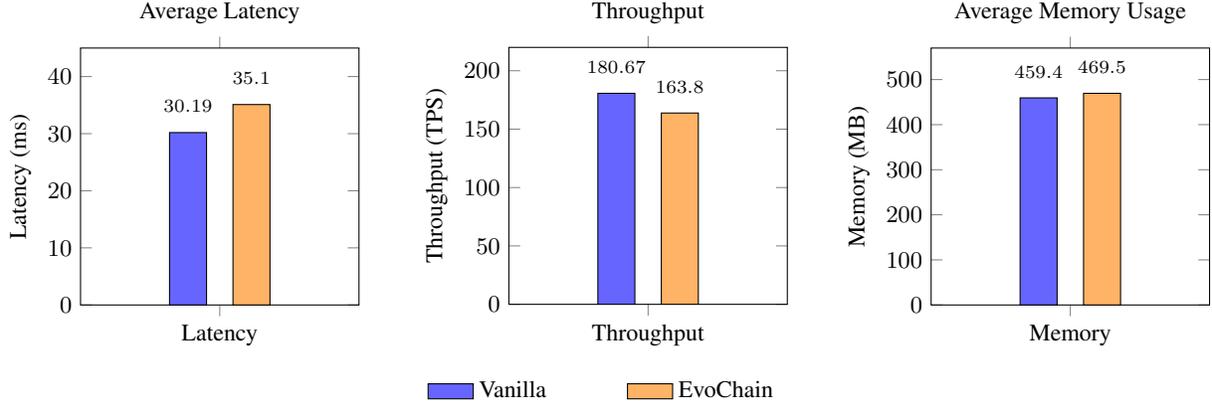

The results are presented in Figure~\ref{fig:winetracker_comparison}.

\subsubsection{Canceling Transactions}

This second test case (TC2) aims to allow us to make a critical evaluation considering the performance of cancel transactions and their impact within the scope of core functionalities.
The performance of CTs was evaluated through incremental rounds of issuing sequential core functionalities. In our implementation, canceling a pending \textit{createGrapes} transaction recursively cancels every subsequent transaction associated with that batchId. In this benchmark, we evaluated the performance of canceling a createGrapes transaction considering the number of dependent transactions, as well as the performance of issuing identical transactions affecting the same keys.

\begin{figure}[!htb]
    \centering
    \small
    \begin{tikzpicture}
        \begin{axis}[
            ybar,
            symbolic x coords={createGrapes, sellGrapes, transformGrapes, cancelCreateGrapes},
            xtick=data,
            xlabel={Function},
            ylabel={Throughput (TPS)},
            legend entries={Round 1, Round 2, Round 3, Round 4, Round 5},
            legend style={at={(0.5,1.15)}, anchor=south, legend columns=3, font=\footnotesize},
            ymin=0, ymax=300,
            x tick label style={rotate=30,anchor=east, font=\small},
            width=0.65\textwidth,
            height=0.4\textwidth,
            enlarge x limits=0.18,
            bar width=8pt
        ]
        
        \addplot+[fill=blue!60] table[x=Name, y=Throughput] {Data/cancel1_evo.dat};
        \addplot+[fill=red!60] table[x=Name, y=Throughput] {Data/cancel2_evo.dat};
        \addplot+[fill=green!60] table[x=Name, y=Throughput] {Data/cancel3_evo.dat};
        \addplot+[fill=orange!60] table[x=Name, y=Throughput] {Data/cancel4_evo.dat};
        \addplot+[fill=purple!60] table[x=Name, y=Throughput] {Data/cancel5_evo.dat};

        \end{axis}
    \end{tikzpicture}
    \caption{Throughput (TPS) by Function and Test Round for TC2 Scenario}
    \label{fig:tc2_throughput_round}
\end{figure}
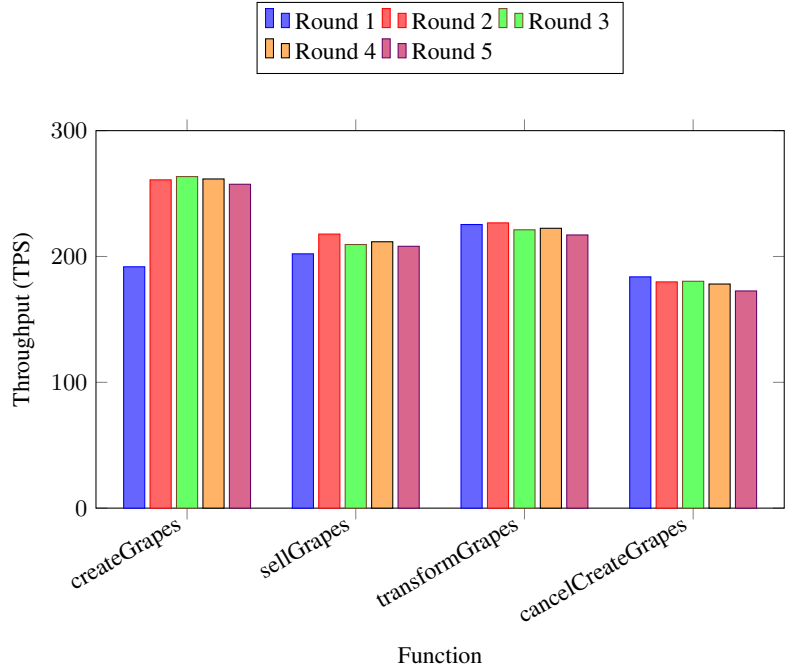

This benchmark consisted of five transactional rounds: creating, selling, transforming, and finally canceling the issued transactions. Ten (10) workers were involved, simulating 10,000 transactions for each round, where each round's rate control started at 400 TPS and gradually increased to 1200 TPS.
Performance results are presented in Figure~\ref{fig:tc2_throughput_round}

\

\subsubsection{Query Transactions}
\label{sec:query-eval}

We evaluate query transactions, considering that these are responsible for consolidating pending transactions that have an expired delay.
This allows us to evaluate the impact on the performance of operations that consolidate other transactions.

To evaluate the impact of transactions that consolidate other transactions by comparing the performance of query transactions when the delay or the queried transaction has and has not already expired. To evaluate the overhead in the system, in the third test case (TC3) we performed 5000 \textit{create}, \textit{sellGrapes}, \textit{transform} and \textit{sellBulk}  transactions, followed by 5000 queries to each of them before the expiring delay and 5000 queries after the expired delay. Involving 10 workers, simulating 5000 transactions for each round, at a rate control between 400 TPS and 1200 TPS. The two result tables are presented on \ref{tab:tc3_consolation} and \ref{tab:tc3_no_consolation}. 

\begin{table}[!htb]
    \centering
    \small
    \caption{Performance of TC3 Consolation Query Operations}
    \label{tab:tc3_consolation}
    \begin{tabular}{lcccc}
        \toprule
        \textbf{Operation} & \textbf{Send Rate (TPS)} & \textbf{Avg. Latency (s)} & \textbf{Throughput (TPS)} \\
        \midrule
        getCreate & 373.1 & 4.39 & 257.9 \\
        getSell & 388.2 & 4.14 & 267.4 \\
        getTransform & 402.0 & 4.31 & 270.5 \\
        getSellBulk & 402.4 & 4.94 & 261.6 \\
        \bottomrule
    \end{tabular}
\end{table}

\begin{table}[!htb]
    \centering
    \small
    \caption{Performance of TC3 No Consolation Query Operations}
    \label{tab:tc3_no_consolation}
    \begin{tabular}{lccc}
        \toprule
        \textbf{Operation} & \textbf{Send Rate (TPS)} & \textbf{Avg. Latency (s)} & \textbf{Throughput (TPS)} \\
        \midrule
        getCreate & 403.8 & 3.07 & 294.6 \\
        getSell & 402.7 & 3.55 & 287.7 \\
        getTransform & 407.4 & 3.25 & 292.7 \\
        getSellBulk & 407.3 & 3.47 & 288.9 \\
        \bottomrule
    \end{tabular}
\end{table}

\subsection{Discussion}

Regarding \textbf{TC1}, where the core functionalities of both versions were compared, and taking into consideration the results presented in \ref{fig:winetracker_comparison}, we can see an average latency of 30.19s from the standard compared to an average of 35.61s on the modified version, representing a 17.95\% increase in latency per transaction.
Throughput average shows a decrease in the modified version, varying from 180.67 TPS to 163.83 TPS, representing a 9.32\% decrease. Average memory also increases from 459.39 MB to 469.54 MB, representing, on average, 2.21\% more usage of memory.

These results are explained by the increase of the transaction size, since more data is now stored within the blocks, as well as in the state database of each peer. The view generation algorithm also increases the code complexity within the peers since more calculations and queries are performed on the state database.

Considering \textbf{TC2}, the lower throughput is immediately noticeable in the first function of the first round. This lower value can be associated with a ``cold start'' from the benchmarking tool and the network infrastructure.
We consider this initial lower throughput in the benchmarking process, but we do not consider it an indicator of long-term performance.
CTs present a lower throughput since they alter the state of three previously issued transactions. This process involves additional computation and validation steps, resulting in a lower transaction processing rate.
It is expected that CTs present lower values of throughput if they recursively cancel a higher number of MTs.
In the long term, each canceled transaction results in additional entries on the blockchain, increasing its overall height. This growing height ends up requiring more storage capacity and more computation resources to query the blockchain. This also applies to the mutable transactions issued, since they carry more information than a standard operation.
The rise in blockchain height increases the confirmation time per block, influencing the endorsement rate and consequently forcing a lower throughput.

By analyzing the results of this benchmark, we can conclude that the CTs do not directly influence 
core transactions.
On more complex transactional systems with larger objects, the view generation algorithm performance can take a hit, especially when the canceled transaction affects the valid state of objects inside consoled transactions, introducing delays to the system.
In our system, transaction validity is only altered when that same transaction is queried, either during the view generation process, before the execution of cancel transactions, or on query transactions.
\textbf{TC3} gives insight into the overhead of these state changes and how the performance of the system can be affected by them.
In cases of forced consolidation by condition, we expect similar results since the process is similar.
The query transactions that do not alter the state of MTs averaged 3.34 s against 4.46 s of the ones that consolidate transactions.
This data shows an increase of 33.53\% in latency.
The throughput went from 290.98 TPS to 264.35 TPS, an 9.15\% decrease. 
This overhead in performance should be expected during runtime, since transactions will be confirmed when they are queried and their mutation policy has expired.

Several security properties of blockchain systems have already been identified and analyzed with regard to their application and challenges in redaction mechanisms \cite{zhang_security_privacy_properties_2019, zhang_exploring_2021}.
We concluded that \appname{} does not influence consistency and tamper-resistance, since our prototype does not alter properties within the Hyperledger Fabric, namely, the Resistance to DDoS since tests on our prototype showed that redaction operations behave, in terms of performance, similarly to core operations, resistance to double-spending since redaction requests work similarly to core functionalities, intrinsic Fabric resistance to double spending is maintained, and pseudonymity, since we do not put any restrictions on its use, has been relegated to Membership Service Providers, if intended.

\section{Conclusion}

In this work, we began with an overview of the core aspects of blockchain systems and how their cryptographic primitives are used to impart desired characteristics, such as decentralization, transparency, immutability, and security. These features are very useful for several applications that require trust, such as financial services, healthcare, supply chain management, among others. 
However, strict immutability of all data can hinder technology adoption in practice, as many systems require the ability to make some changes to fix mistakes or handle effects of malicious intrusions.

We proposed \appname{}, a chaincode development framework that introduces controlled mutability by creating a transaction model that, coupled with access control policies, allows for grace periods during which corrections and recovery can be applied to data within blockchain systems.
A view generation component is responsible for generating a consistent view of the chain, using new fields such as $Submission$ $Time$, $Permanent$ $State$ $Time$, $Validity$ and $Delay$.
Our framework aims to enhance the flexibility regarding the data, allowing for easy restoration of the state of objects that were affected by unwanted operations.

A prototype of a supply chain application was implemented with and without the use of \appname{}, allowing us to perform a comparative analysis of the performance overhead.
The framework version increased latency by around 15\% and reduced throughput by around 10\%.
Canceled transactions contribute to the lower throughput due to extra computation and validation steps, but these will be exceptional cases. In addition, over time, increasing blockchain size requires more storage and computation resources. However, these performance impacts are limited and are balanced by the enhanced functionality and greater control over transactions.

The use of \appname{} as a base for chaincode development introduces a performance overhead to the system. 
In identical conditions, core functionalities present a 9.32\% decrease in throughput, cancel transactions increase the blockchain data volume and influence the view generation complexity and operations that query transactions with a mutation policy expired suffer a 9.15\% decrease in throughput.
Inherent security properties regarding blockchain systems are not compromised since \appname{} implements a transactional model within the chaincode, without altering the base properties of  Hyperledger Fabric.

In conclusion, the ability to fix errors and correct the effects of intrusions while preserving trust-inducing properties can be a differentiator in the decision to adopt blockchain technology in more applications.

\clearpage

\section*{Acknowledgments}

This work was partially supported by national funds through Funda\c{c}\~ao para a Ci\^encia e Tecnologia (FCT) with reference UIDB/50021/2020 (INESC-ID), by the European Union’s Horizon 2020 research and innovation programme under grant agreement No 952226, project BIG (Enhancing the research and innovation potential of Tecnico through Blockchain technologies and design Innovation for social Good), and by Project Blockchain.PT – Decentralize Portugal with Blockchain Agenda, (Project no 51), WP 1: Agriculture and Agri-food, Call no 02/C05-i01.01/2022, funded by the Portuguese Recovery and Resilience Program (PRR), The Portuguese Republic and The European Union (EU) under the framework of Next Generation EU Program.

\bibliographystyle{unsrt}  
\bibliography{paper}

\end{document}